\documentclass[review]{elsarticle}
\usepackage{longtable}
\usepackage{amsmath}
\usepackage{graphicx}
\usepackage{amssymb}
\usepackage{epstopdf}
\usepackage{pdfpages}
\usepackage{dcolumn}
\usepackage{hyperref}
\usepackage{bm}

\newcommand{\be}{\begin{equation}}
\newcommand{\ee}{\end{equation}}
\newcommand{\susu}{SU(2)$\times$SU(2) }

\begin{document}
\begin{frontmatter}

\title{Special Unitary Particle Pusher for Extreme Fields}

\author{D.F. Gordon, B. Hafizi}
\address{Naval Research Laboratory, Plasma Physics Division, Washington, DC 20375, USA}

\begin{abstract}
An exact momentum update useful for particle codes was previously given.  The expressions involved were unwieldy.  By treating the momentum as a representation of \susu instead of SO(3,1), a more compact expression for the exact momentum update is obtained.  An expansion in powers of the timestep can be formulated such that invariance properties are exactly preserved, and push rates comparable to the standard Boris pusher are obtained.
\end{abstract}


\end{frontmatter}

\section{Introduction}

Electromagnetic particle-in-cell codes utilize a variety of algorithms to update the particle momentum \cite{boris70,verboncoeur05,vay2008,higuera2017,gordon17.AAC,gordon2017,petri2020.arxiv}. The accuracy of the standard Boris pusher \cite{boris70,qin2013} is reduced when $eA/mc^2 \gg 1$, where $A$ is the vector potential \cite{arefiev2015}.  Reference~\cite{gordon17.AAC} gives an accuracy condition on the timestep of any Boris-factorized pusher as
\be
\label{eq:splitting-condition}
\left(\frac{eF\Delta s}{2mc}\right)^3 \ll 1.
\ee
where $\Delta s$ is the proper time step, and $F$ is a typical field tensor magnitude.  This condition is a direct consequence of the operator splitting scheme, namely, the separation of the time translation operator into two factors involving only the electric field, and a third factor involving only the magnetic field.  Once this factorization is adopted, the condition (\ref{eq:splitting-condition}) persists, even if the individual factors (which become additive terms in certain approximations) are evaluated exactly.  The fundamental source of the error is that boost operators and rotation operators do not commute.

References~\cite{vay2008,higuera2017} retain the Boris-type factorization, but modify the temporal discretization in order to better capture certain desirable characteristics of the particle orbits, such as accurate cancellation of electric and magnetic field forces, and volume preservation.

References~\cite{gordon17.AAC,gordon2017} give a closed form expression for the time translation operator that is exact in a uniform field, and therefore eliminates the condition on the time step (\ref{eq:splitting-condition}).  The primary disadvantage of the expression in \cite{gordon17.AAC,gordon2017} is that it is unwieldy.  Reference~\cite{petri2020.arxiv} gives another closed form expression for the time translation operator that is exact in a uniform field, but simplifies the expression by transforming to a special Lorentz frame.

The expressions from \cite{gordon17.AAC,gordon2017,petri2020.arxiv} operate on four-vectors, which are representations of the group of transformations known as SO(3,1).  As is well known, the group \susu is isomorphic to SO(3,1).  It turns out that treating particle momentum as a representation of \susu results in a much more compact time translation operator.  For brevity, we sometimes use ``unitary'' in place of ``special unitary.''

\section{Special Unitary Time Translation}

This section derives the unitary form of the exact time translation operator of the four-momentum in a constant, uniform, electromagnetic field, with arbitrary polarization.   Ordinarily one has a four-velocity $u^\mu$, satisfying the equation of motion $mdu^\mu /ds = eF^{\hspace{0.05in}\mu}_\nu u^\nu$, where $s$ is the proper time.  For constant, uniform fields, the exact solution is furnished by taking the matrix exponential.  The expression for the matrix exponential is tractable, but somewhat onerous, even after employing a rotation to simplify it \cite{gordon17.AAC}.  The particle pusher obtained this way is exactly Lorentz invariant.  Preservation of the Euclidean norm of the velocity in a magnetic field is merely a special case.

In order to obtain an equivalent pusher that can be expressed more easily, we employ the well known relationship between four-vectors and second rank spinors \cite{Landau-qed}.  Let $\zeta$ be the $2\times 2$ matrix representing the spinor that represents momentum.  Then
\be
\label{eq:u}
u^\mu = \frac{1}{2}{\rm tr} (\sigma^\mu \zeta)
\ee
Here $\sigma^\mu$ are the Pauli matrices if $\mu=[1,2,3]$ and the identity if $\mu=0$.  The inverse operation gives the matrix
\be
\label{eq:zeta}
\zeta = \left(
\begin{matrix}
u^0+u^3 & u^1 - iu^2 \\
u^1 + iu^2 & u^0-u^3
\end{matrix}
\right)
\ee
The steps used in \cite{Landau-qed} to derive the boost and rotation operators are a useful guide in deriving the operator of time translation.  Let the time translation operator be denoted $\Lambda(\Delta s)$, where $\Delta s$ is any interval in proper time.  The spinor is transformed as
\be
\label{eq:advance}
\zeta(s+\Delta s) = \Lambda(\Delta s)\zeta(s)\Lambda^\dag(\Delta s)
\ee
It is useful to define the generator $\lambda$ by $\Lambda(ds) = 1 + \lambda ds$.
The equation of motion for $\zeta$ is then
\be
\frac{d\zeta}{ds} = \lambda\zeta + \zeta\lambda^\dag.
\ee
Using the equations of motion for $u$ and $\zeta$, along with (\ref{eq:u}), one obtains
\be
\lambda = \bm{\sigma} \cdot \bm{\Omega}
\ee
where bold type is used for three-vectors, and
\be
\bm{\Omega} = \frac{1}{2}\frac{e}{mc}\left({\bf E} + i {\bf B}\right)
\ee
The full time translation operator is again furnished by a matrix exponential, $\exp(\lambda\Delta s)$.  Thanks to the anticommuting properties of the Pauli matrices this can be reduced to the practical expression
\be
\label{eq:exact}
\Lambda(\Delta s) = \cosh\Omega\Delta s + \bm{\sigma}\cdot \bm{\omega} \sinh\Omega\Delta s
\ee
where
\be
\Omega = \sqrt{\bm{\Omega}\cdot\bm{\Omega}}
\ee
and
\be
\bm{\omega} = \frac{\bm{\Omega}}{\Omega}
\ee
Note that $\Lambda(\Delta s)$ is an even function of $\Omega$, so that the sign of the square root is immaterial.  Hereinafter, the pusher algorithm that uses Eq.~(\ref{eq:exact}) is referred to as the exact-unitary pusher.

\section{Invariants}

The familiar invariants associated with charged particle motion take a new form in the \susu representation of the momentum.  It is convenient to define
\be
\bm{\Psi} = \bm{\Omega}\Delta s,
\ee
which can be thought of as a set of Minkowski-type angles.  In a pure electric field, the magnitude $\Psi = \sqrt{\bm{\Psi}\cdot\bm{\Psi}}$ is real, and the hyperbolic functions have real arguments, leading to a Lorentz boost.  In a pure magnetic field, $\Psi$ is imaginary, and the hyperbolic functions become equivalent to trigonometric functions of a real number, leading to a rotation.  In the case where ${\bf E}\cdot{\bf B} = E^2-B^2 = 0$ (i.e., plane wave fields), $\bm{\Psi}\cdot\bm{\Psi} = 0$ and the time translation operator reduces to $\Lambda = 1 + \bm{\sigma}\cdot\bm{\Psi}$.

In the \susu representation, Lorentz invariance corresponds to the invariance of the determinant, i.e.,
\be
\frac{d}{ds}{\rm det}\zeta = 0.
\ee
It is straightforward to demonstrate that the transformation $\Lambda(s)\zeta(0)\Lambda(s)^\dag$ satisfies this condition exactly.  If ${\bf E}=0$ is inserted into the transformation, then the trace is also invariant, i.e.,
\be
\frac{d}{ds}{\rm tr}\zeta = 0.
\ee
This corresponds to conservation of energy in a pure magnetic field.  In the case ${\bf E}\cdot{\bf B} = E^2-B^2 = 0$,
\be
\frac{d}{ds}{\rm tr}\left(1-\bm{\sigma}\cdot {\bf e}_\parallel\right)\zeta = 0
\ee
where ${\bf e}_\parallel$ is a unit vector in the direction of ${\bf E}\times{\bf B}$.  This corresponds to the invariance of $k^\mu u_\mu$ in a plane wave with wavevector $k^\mu$.

\section{Invariant Truncated Expansion}

It is possible to form a truncated expansion in the timestep that exactly preserves all the invariance properties mentioned above.  This expansion is useful because it eliminates special functions and square roots entirely from the formulation, which improves performance on typical computer hardware.  The time translation operator, expressed as a limit, is
\be
\Lambda = \lim_{n\rightarrow\infty} \left(1 + \bm{\sigma}\cdot\bm{\Psi}/n\right)^n
\ee
The obvious truncation, i.e., simply taking $n$ as finite, does not preserve all the invariants.  An approximation which does preserve all the invariants is
\be
\label{eq:trunc:order:n}
\Lambda^{(n)} = \left(1 + \bm{\sigma}\cdot\bm{\Psi}/n\right)^{n/2} \left(1 - \bm{\sigma}\cdot\bm{\Psi}/n\right)^{-n/2}
\ee
This has unit determinant, as can be seen by noting that
\be
\det(1 + \bm{\sigma}\cdot\bm{\Psi}/n) = \det(1 - \bm{\sigma}\cdot\bm{\Psi}/n)
\ee 
and using the well known properties of the determinant.  Since all the matrix factors involved in the time translation have unit determinant, invariance of the Minkowski norm is unaffected by the truncation (\ref{eq:trunc:order:n}).  In a pure magnetic field, the energy is also exactly invariant, as can be verified by direct evaluation of the trace with ${\bf E} = 0$. Finally, in a plane wave, the exact and truncated operators are identical due to $\bm{\Psi}\cdot\bm{\Psi} = 0$. Therefore the invariance of $k^\mu u_\mu$ is also preserved.

As is shown below, the second order time translation operator $\Lambda^{(2)}$ is extremely accurate in practice.  Making use of $(\bm{\sigma}\cdot\bm{\Psi})^2 = \bm{\Psi}\cdot\bm{\Psi}$ gives the practical expression
\be
\label{eq:trunc}
\Lambda^{(2)} = \frac{1 + \bm{\sigma}\cdot\bm{\Psi} + \bm{\Psi}\cdot\bm{\Psi}/4}{1-\bm{\Psi}\cdot\bm{\Psi}/4}
\ee
An immediate optimization is to factor out the denominator from both $\Lambda^{(2)}$ and its Hermitian conjugate.  The two denominators taken together amount to division of the entire result by a real number.  Hereinafter, the pusher algorithm that uses Eq.~(\ref{eq:trunc}) is referred to as the quadratic-unitary pusher.

\section{Unitary Pusher Algorithm}

The implementation of the above expressions in a numerical particle pusher is straightforward.  One has essentially three steps: (i) form the spinor $\zeta$ using (\ref{eq:zeta}) , (ii) advance the spinor using (\ref{eq:advance}), and (iii) restore the four-velocity using (\ref{eq:u}).  These steps involve $2\times 2$ matrix manipulations with complex numbers, which presents little difficulty.  If the expansion (\ref{eq:trunc}) is used, there are no caveats.  If the exact operator is used, one must take precautions in field free regions, where the vector $\bm{\omega}$ is not well defined numerically.  In practice this is easy to manage.  For example, using the approximation
\be
\bm{\omega} \approx \frac{\bm{\Omega}}{\Omega + N^{-1}\Delta s^{-1}}
\ee
with $N\gg 1$ is sufficient in practice.  If one is concerned about the possibility of cancellation in the denominator, one can choose the sign of the square root in the definition of $\Omega$ such that the real part is always positive.

Generalization to the case of non-uniform fields is carried out in the usual way.  In the following sans-serif type represents an abstract four-vector.  The spacetime dependence of the field distribution is accounted for by successively updating the world point of the particle, ${\sf x}$, and using the composition ${\sf F}(s) = {\sf F}[{\sf x}(s)]$.    The world line is discretized by
\be
{\sf x}(s+\Delta s) = {\sf x}(s) + {\sf u}\Delta s,
\ee
where ${\sf x}$ and ${\sf u}$ are leapfrogged in time.  The step size $\Delta s$ must be sufficiently small to resolve variations in the field.

For single particle trajectories, advancing in proper time raises no problems, and in fact, is often advantageous \cite{gordon17.AAC}.  For self-consistent simulations, one usually has to advance all particles through an interval of time as measured by a laboratory frame clock, and therefore the proper time advance of every particle is different.  The proper time step $\Delta s$ for any particle is formally obtained by integrating
\be
\frac{dt}{ds} = \frac{1}{2}{\rm tr}\left[\Lambda(s)\zeta(0)\Lambda^\dag(s)\right]
\ee
and inverting the resulting expression $t(s)$.  Here, $t \equiv x^0/c$.  Carrying out the integration to second order in $s$, and the inversion to second order in $t$, results in
\be
\label{eq:timestep}
\Delta s^{(2)} = \frac{\Delta t}{\gamma}\left(1-\frac{e{\bf E}\cdot\bm{\beta}}{2\gamma mc}\Delta t\right)
\ee
where $\bm{\beta} \equiv {\bf u}/u^0$ and $\gamma \equiv u^0$.  Calculation of the proper time step in the context of SO(3,1) is discussed in \cite{petri2020.arxiv}.

\section{Testing}

The Naval Research Laboratory TRACKER code implements the Boris \cite{boris70}, Vay \cite{vay2008}, Higuera-Cary (HC) \cite{higuera2017}, quadratic-unitary, exact-unitary, and SO(3,1) pushers.  The advantages of the SO(3,1) pusher, which is mathematically equivalent to the exact-unitary pusher, are discussed in \cite{gordon17.AAC} and in the supplementary material of \cite{gordon2017} (the term ``covariant pusher'' is used in these references). These articles show that the SO(3,1) pusher gives the correct solution in orders of magnitude fewer steps than the standard Boris pusher.  The exact-unitary pusher, being mathematically equivalent, has all the same properties, to within a round-off error.

\begin{figure}
\includegraphics[width=3in]{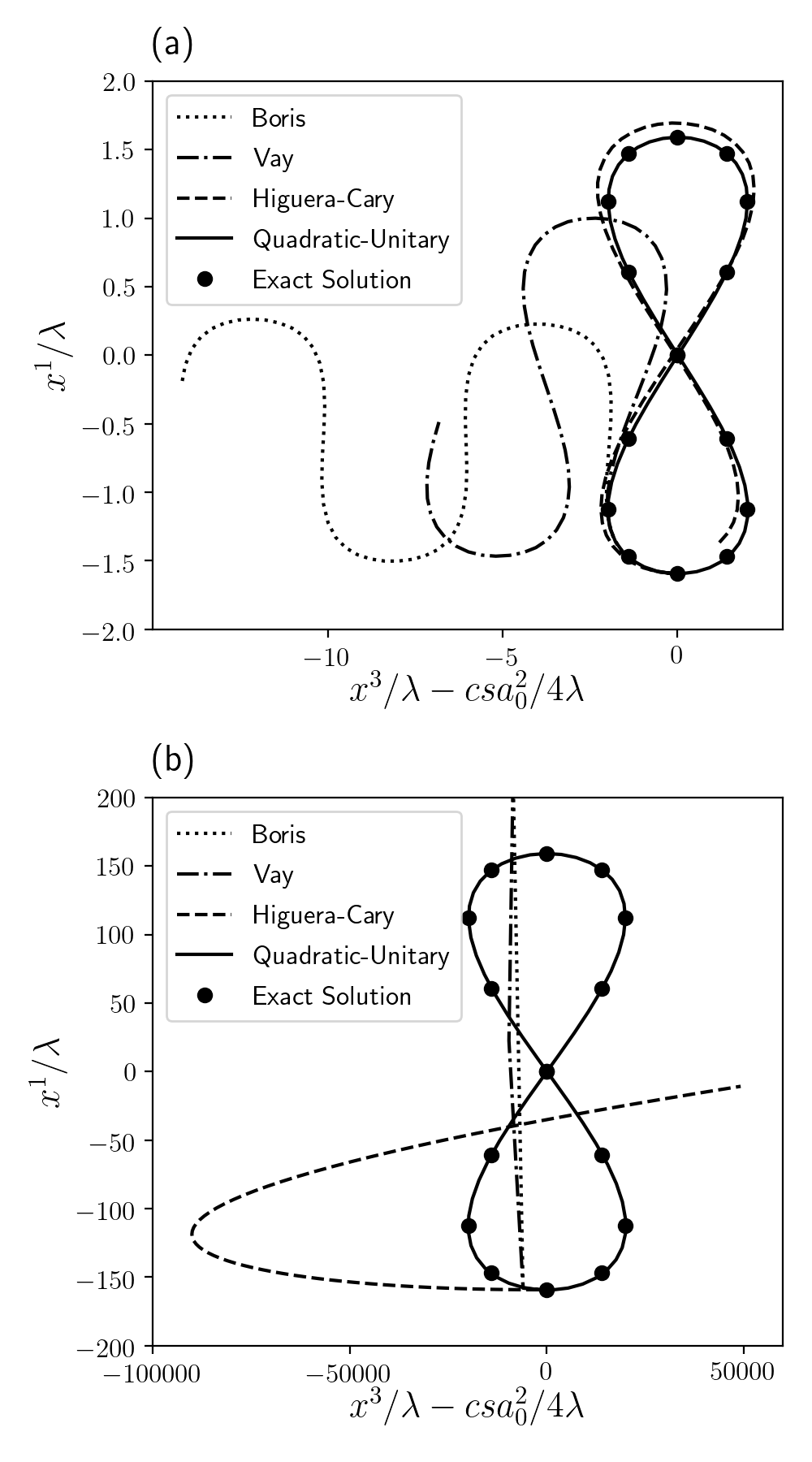}
\caption{Orbits in a plane wave as computed by several numerical particle pushers, and the analytical solution, for (a) $a_0 = 10$ and (b) $a_0 = 1000$.  The orbits are displayed in a Galilean frame comoving with the particle's true orbit.  The proper time step is fixed to the same value in all cases.}
\label{fig:orbits}
\end{figure}

It turns out the quadratic-unitary pusher (\ref{eq:trunc}) gives results very close to the exact-unitary and SO(3,1) pushers.  As an illustration, Fig.~\ref{fig:orbits} shows the spatial orbit of a single particle in a plane wave, as computed by the Boris, Vay, HC, and quadratic-unitary pushers.   The analytical solution at several points is also shown for comparison.  The wave amplitude is $a_0 = 10$ in panel (a), and $a_0 = 1000$ in panel (b).  The radiation wavelength, $\lambda$, can be scaled arbitrarily.  In all cases, the calculation is run for 63 steps, with $c\Delta s = 0.1\lambda/2\pi$. Note that a constant step in proper time naturally leads to a constant phase step \cite{gordon17.AAC}.  For the lab frame pushers, the inverse of (\ref{eq:timestep}) is used to obtain the lab frame step $\Delta t$.    For $a_0 = 10$, the quadratic-unitary solution is indistinguishable from the analytical solution.  The HC pusher gives nearly the right behavior, but the errors are visible.  The Boris and Vay pushers are not viable for the given parameters.  For $a_0 = 1000$, out of the four numerical solutions, only the quadratic-unitary pusher is viable.  We verified that the exact-unitary and SO(3,1) pushers are also effective in this case.

\begin{figure}
\includegraphics[width=3in]{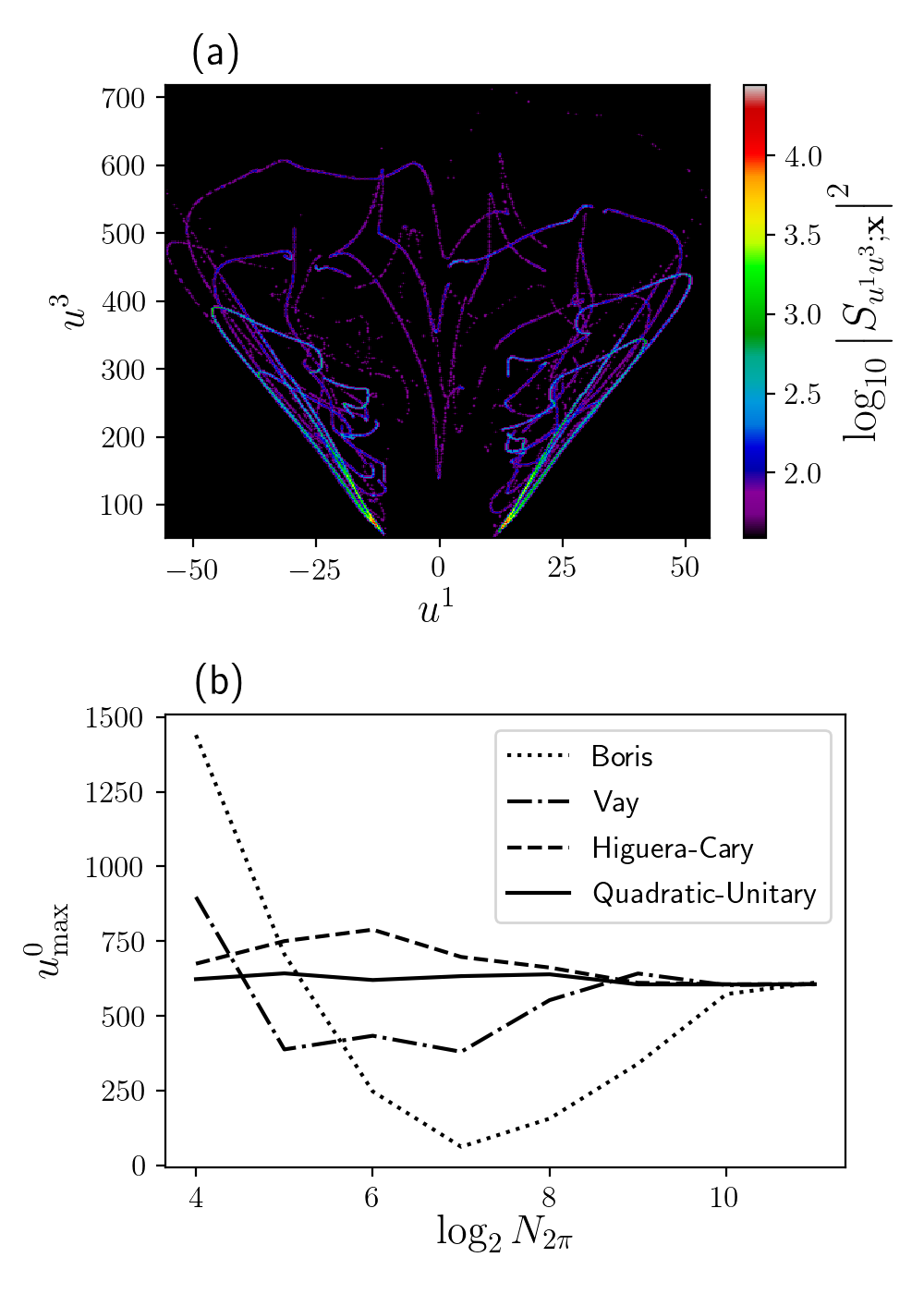}
\caption{Test case of tunneling ionization of an Ar$^{17+}$ ion positioned at ${\bf x} = 0$.  Panel (a) shows the momentum distribution as computed by the quadratic-unitary pusher.  The color scale represents the modulus squared of the quasi-classical S-matrix of the interaction, i.e., the probability to produce a given photoelectron momentum ${\bf u}$ from an ion at position ${\bf x}$.  See Ref.~\cite{gordon2017} for details.  Panel (b) shows the rate of convergence of the several pushers, in terms of the maximum photoelectron energy, $u^0_{\rm max}$, and the number of steps per optical period, $N_{2\pi}$.}
\label{fig:u1u3}
\end{figure}

As a more elaborate test, consider photoionization of hydrogen-like argon, under illumination by a 10 petawatt class laser pulse, following Ref.~\cite{gordon2017}.  The momentum distribution as computed by the quadratic-unitary pusher is displayed in Fig.~\ref{fig:u1u3}(a).  This reproduces Fig.~3(a) from \cite{gordon2017}.  The peak vector potential is $100mc^2/e$, the wavelength is 0.8 $\mu$m, the spot size is 5 $\mu$m, and the pulse duration is 30 fs.  In order to test the convergence of the various pushers in this scenario, the maximum photoelectron energy is plotted vs. the number of steps per optical cycle, $N_{2\pi}$, in Fig.~\ref{fig:u1u3}(b).  Each point of each curve represents $10^3$ trials, where the ionization phase is the variable parameter.  The random number generator is seeded with the same value for each set of $10^3$ trials.  The quadratic-unitary, exact-unitary, and SO(3,1) results are almost indistinguishable, so only the quadratic-unitary curve is shown.  In this scenario, the quadratic-unitary pusher converges fastest, and the Boris pusher converges slowest.  In all cases the time step is constant in proper time.

\section{Performance}

\begin{figure}
\includegraphics[width=3in]{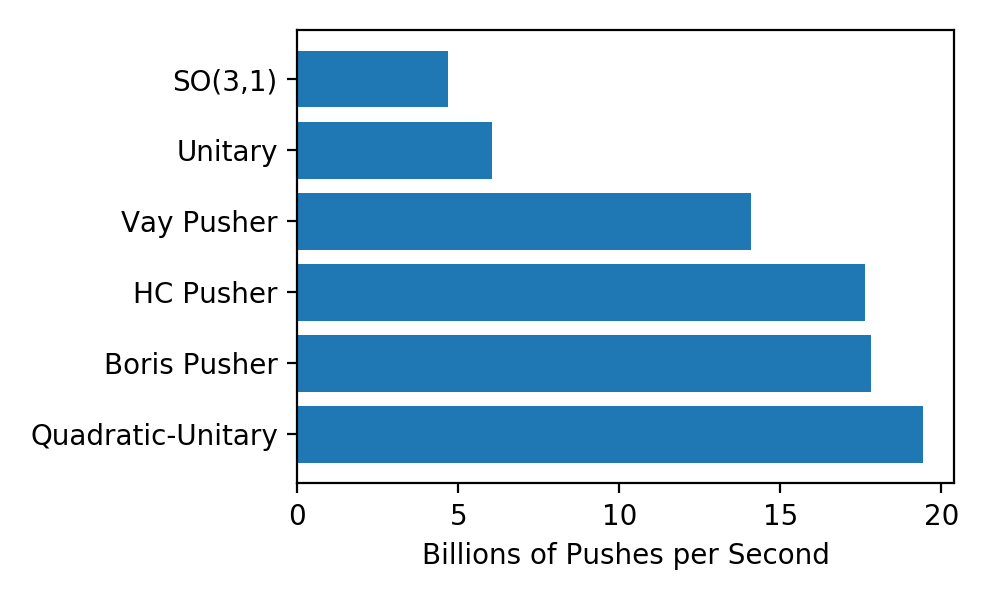}
\caption{Push rates on an NVIDIA Titan V GPGPU.}
\label{fig:perf}
\end{figure}

The TRACKER code is optimized for parallel processing on either central processing units (CPU) or general purpose graphical processing units (GPGPU).  To test the ``absolute performance'' of the various pushers, defined as particles pushed per second on a single device, motion in a plane wave field of $10^5$ particles for $10^6$ steps is computed.    It must be emphasized that this measure of performance is decoupled from the error tolerance.  The device is an NVIDIA Titan V GPGPU.  The programming model is Python with PyOpenCL acceleration.  Extraneous algorithms in TRACKER, such as the ionization algorithm, adaptive time stepping, and completion logic, are turned off.  The cost of the field evaluation is minimized by using plane wave formulas.  The results are summarized in Fig.~\ref{fig:perf}.  The quadratic-unitary pusher is the fastest, with the HC and Boris pushers close behind.  The exact unitary and SO(3,1) pushers, which keep all orders in $\Delta s$, are much slower.  Interestingly, the performance ranking in Fig. 3 is consistent with the assumption that the number of special function evaluations (including square roots) dominates the computing time.  It should be acknowledged that the rankings in Fig.~\ref{fig:perf} might change after a determined optimization effort with any given pusher.

\section{Conclusions}

Previous work describes an extreme field particle pusher operating in the representation of SO(3,1).  This pusher is effective, but the expressions involved are onerous to program.  The expressions presented herein are much simpler, yet mathematically identical.  This is achieved by operating in the representation of \susu instead of SO(3,1).  Furthermore, an expansion in the time step can be carried out to any desired order, while maintaining the exact invariance properties of the pusher.  The second order expansion is accurate, easy to program, and involves no special functions or square roots.  It gives push rates comparable to the standard Boris pusher, and produces accurate orbits in extreme fields, in orders of magnitude fewer steps.

\section{Acknowledgments}
This work was supported by the U.S. Department of Energy Interagency Agreement 89243018SSC000006.


\end{document}